**Superconducting anti-dot arrays from self-assembly template methods**

A.A.Zhukov[a], E.T.Filby[a], M.A.Ghanem[b], P.N.Bartlett[b] and P. A. J. de Groot[a]

[a]*School of Physics and Astronomy, University of Southampton, SO17 1BJ, UK,*

[b]*School of Chemistry, University of Southampton, Southampton, SO17 1BJ, UK*

**Abstract**

We present results of electrochemical deposition of superconducting Pb in the pores of templates prepared by self-assembly from colloidal suspensions of polystyrene latex spheres. This technique enables us to create highly ordered superconducting nano-structures with 3D architectures on length scales ranging from 50 – 1000 nm. The prepared samples show pronounced Little-Parks oscillations in $T_c$ and matching effects in magnetization and magnetic susceptibility. Real and imaginary parts of susceptibility follow a universal Cole-Cole curve. Self-field effects play an important role in commensurability behaviour of magnetic moment at low temperatures.

## 1. Introduction

Superconducting nano-structures demonstrate large potential for both fundamental research [1-3] and electronic applications [4-6]. Combining thin film growth techniques with sub-micron lithography is a standard method for the preparation of such materials. However for lateral dimensions below 100nm-200nm relatively cheap and fast optical lithography techniques become inapplicable. The cost of ion or electron lithography is uneconomically high for industrial applications. Alternative routes are needed for creating highly-ordered structures on length scales down to a few nanometres. In this paper we present results for superconducting nanostructured samples prepared by electrochemical deposition through different templates produced by self-assembly of polystyrene spheres. These well ordered structures demonstrate pronounced commensurability effects in magnetization and magnetic susceptibility.

## 2. Sample preparation and experimental techniques

The substrates have been prepared by evaporating a 10 nm chromium adhesion layer, followed by 300 nm of copper, onto 1 mm thick glass microscope slides. The copper substrate was cleaned by sonication in propanol for 1 h followed by rinsing with deionised water. The polystyrene sphere templates were assembled by sticking two copper substrates (1.5 x 2.0 cm$^2$, modified with thiol) together using two layers of Para-film as a spacer (about 150 μm thick).

The space between the two substrates was filled with about 40 µL sphere solution and the sample was kept vertically inside an incubator at 30 °C for 24 h. After drying, the template appears opalescent, as expected, with colors from blue to red clearly visible. The resulting templates are robust and adhere well to the copper substrates. More details of the template preparation method are described in Refs.7 - 9. The electrochemical deposition was carried out using an EG&G283 potentiostat and a conventional three-electrode glass cell (15 ml volume capacity). Large area (1 cm$^2$) platinum gauze was used as counter electrode and a home made saturated calomel electrode (SCE) as the reference electrode. Lead film electrodeposition was carried out from a plating solution (50 mM lead carbonate and 1.0 M methane sulphonic acid) at constant potential of -0.49 V vs. SCE and deposition charge density of about 1.0 C cm$^{-2}$ µm$^{-1}$. After the lead electrodeposition was complete, the polystyrene sphere templates were removed by soaking the films in tetrahydrofuran (THF) for 2h.

Fig.1 presents a scanning electron microscope (SEM) image of a Pb film deposited through a template of spheres with a diameter of 700nm. The crystal structure, composition and morphology of the films were characterised using SEM, energy dispersive X-ray spectroscopy and X-ray diffraction.

In this work we present results for four different crystals prepared using 700nm spheres (#700a – thickness ~175nm, #700b – thickness ~350nm, #700c – thickness 700nm) and 900nm spheres (#900 – thickness ~450nm). Magnetic measurements have been performed using a Hall probe (sensitive area 50µm×50µm) magnetometer [10]. The ac-susceptibility has been determined from the Hall probe response to ac-magnetic field with a frequency of 62Hz and amplitude in the range 1μT - 1mT.

## 3. Experimental results and discussion

The nanostructured films demonstrate a sharp superconducting transition. The Fig.2 shows the ac-susceptibility ( $H_{ac}$ = 0.15 Oe) of a nanostructured Pb film prepared using 900nm spheres (sample #900). The onset of superconducting transition is $T_c$ = 7.02K in good agreement with the transition temperature for the plain film of 7.3K. The width of superconducting transition (10-90% levels) is 0.15K. Other samples show the value of $T_c$: 6.35K, 6.64K and 7.28K and width 1.3K, 0.18K and 0.23K for #700a, #700b and #700c, respectively.

As can be seen from Fig.3 the studied films demonstrate a strong Little-Parks effect [11]. The superconducing transition temperature shows pronounced oscillations with the magnetic field. Commensurability results in sharp peaks at $H_n = n\Phi_o/S$ with n = 0, ±1, ±2, …, $\Phi_o$ is the flux quantum and $S = 3^{1/2}d^2 / 2$ ( d is the period of the hexagonal structure). The peaks broaden with n. As can be seen from Fig.3 the amplitude of oscillations exceeds 1% of $T_c$. This is comparable to result for lithographically prepared structures [12]. From the period of such commensurability oscillations in $T_c$ and other characteristics we can find the corresponding d values. The periods are 40.9, 40.6, 36.4 and 26.2Oe for samples #700a,b,c and #900 respectively. The corresponding d values of 765, 767, 810 and 956 are close to the sphere diameters, as expected.

The magnetization loops show rather complicated behaviour depending on temperature. Fig.4 demonstrates magnetization curves for the film prepared using 700nm spheres with thickness 350nm (#700b). At low temperatures T < 5.5K the inhomogeneous self-field created by shielding currents inside the sample and flux jumps suppress the commensurability. For a disc the self-field value reaches a maximum in the centre $H_p = j_c \log(L/t)/2$ ($j_c$ - critical current density, L –transverse size, t – thickness of the film) [13]. In our experiment we characterise the self-field effect by a switching field, which is required to change the direction of the current in the critical state. As can be seen from the inset to Fig.4 the commensurability oscillations vanishes when the self-field approaches the period of oscillations. The position of the peaks is also affected by the self-field. They shift to negative or positive directions for increasing or decreasing branches of the magnetization curve as shown by arrows in Fig.4.

At higher temperatures T > 5.5K commensurability features can be first seen at high fields where $j_c$ is smaller. This corresponds to steps in the critical current. Then commensurability expands to all fields on the decreasing branch of the magnetic fields. Commensurable points up to n=6 can be detected. First no commensurability exists for increasing magnetic field. In this case the commensurability effects start to develop only at 5.7K but until 6.2K the commensurability anomalies remain weaker than for the decreasing branch of the magnetic field. This behaviour suggests the influence of the surface or edge barrier leading to more disorder in the vortex lattice for increasing magnetic field.

For T > 5.8K the steps at low fields start to transform into peaks. At 6.2K this transformation is finished for all fields. At 6.6K the irreversible magnetisation for the sample #700b vanishes.

As can be seen from Fig.5 the ac-susceptibility measurements also reveal commensurability caused by the periodic nano-structure. One of the most unusual features of $\chi$ is the presence of a paramagnetic response at high temperatures. The origin of this anomaly is not yet clear. Neither similarly prepared plain films nor standard lithographic 2D structures [14-16] demonstrate this behaviour. There are significant differences in the real and imaginary parts of susceptibility. The in-phase $\chi'$ susceptibility shows sharp dips at commensurate points as increased critical current gives steep rise to the in-phase shielding. The out-of-plane component $\chi''$ also shows minima at commensurate points for low temperatures and magnetic fields. However, for higher temperatures and fields the situation changes and peaks (i.e.maxima) in $\chi''$ are realised at commensurate points [14]. The transition point between the two different behaviours is determined by the main maximum point of $\chi''$. Further insight can be reached from the Cole-Cole plot of $\chi''$ vs. $\chi'$ [17, 18]. The experimental data from Fig.5 is used for this purpose in Fig.6. We see that the data essentially follow a universal temperature independent curve as in plain films [18].

The ac-susceptibility of homogeneous superconductors has been extensively analysed theoretically [19, 20]. In the framework of Bean model the real and imaginary susceptibilities are expressed in terms of a dimensionless parameter $h = H_{ac} / H_p$. In particular, for a strip $\chi' = -h^{-1} \tanh h$ and $\chi'' = -h^{-1}\tanh(h) + 2h^{-1}\tanh(h/2)$ [18]. This corresponds to a universal curve parametrically determined by h and independent from critical current variations with magnetic field and temperature as observed in our experiment. For temperature decreasing below $T_c$, $H_p$ increases from zero to a value much larger than $H_{ac}$. Then relations above give $\chi'$ monotonically changing from 0 to –1, and $\chi''$ showing a maximum at $h \approx 2.33$. This maximum in $\chi''$ resulting into an arch-like Cole-Cole plot is a universal feature for different sample shapes, current-voltage characteristics and magnetic field dependences of the critical current [19, 20].

Existence of an arch-like curve (Fig.6) explains transformations in $\chi''$ commensurability behaviour. Obviously below the maximum (for temperatures lower 6.3K), $\chi'$ and $\chi''$ oscillate in phase and result in dips for commensurate states. Above the maximum the real part continues to show a dip in accordance with behaviour of the critical current and related $H_p$ value. However, the negative derivative of the Cole-Cole plot results in out-of-phase oscillations with peaks in $\chi''$.

## 4. Conclusions

We have prepared Pb films with periodical arrays of holes using electrochemical deposition of lead in the interstitial spaces of self-assembled templates. The resulting superconducting nanostructures can be prepared with 3D-achitectures on length scales from 20 nm to several µm. The prepared samples demonstrate pronounced Little-Parks oscillations in $T_c$ and many matching anomalies in magnetization and ac-susceptibility. The ac-susceptibility shows different behaviour for the in-phase and out-of-phase components at low temperatures. Both components follow a universal curve in a Cole-Cole plot. Commensurability behaviour of magnetisation is strongly affected by self-field at low temperatures. This results in a shift in peak positions and the gradual vanishing of oscillations with decreasing temperature. The template growth technique offers the potential of a low-cost preparation method for sub-micron patterned superconducting media.

This work has been supported by the EPSRC.

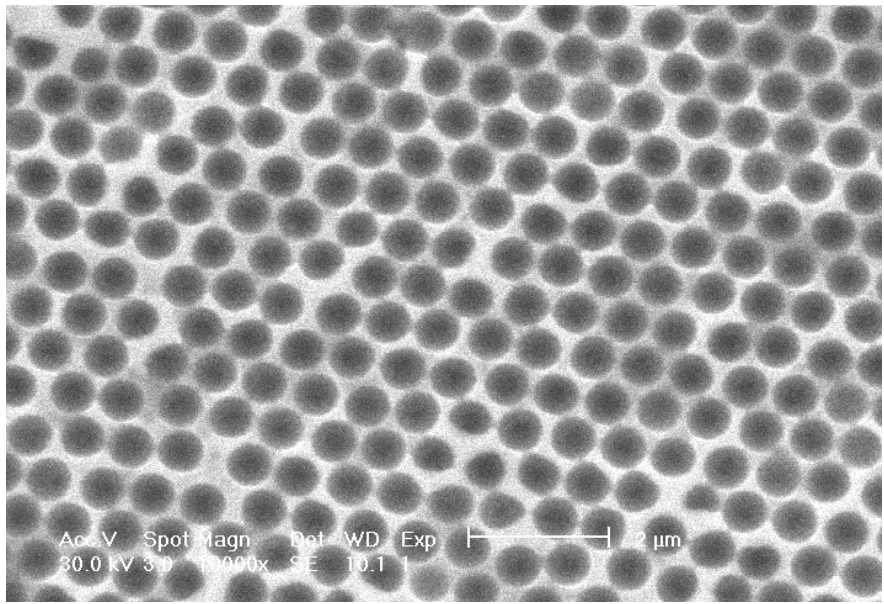

Fig.1 SEM image of Pb film prepared using a template from 700 nm polystyrene spheres. The scale bar is 2μm.

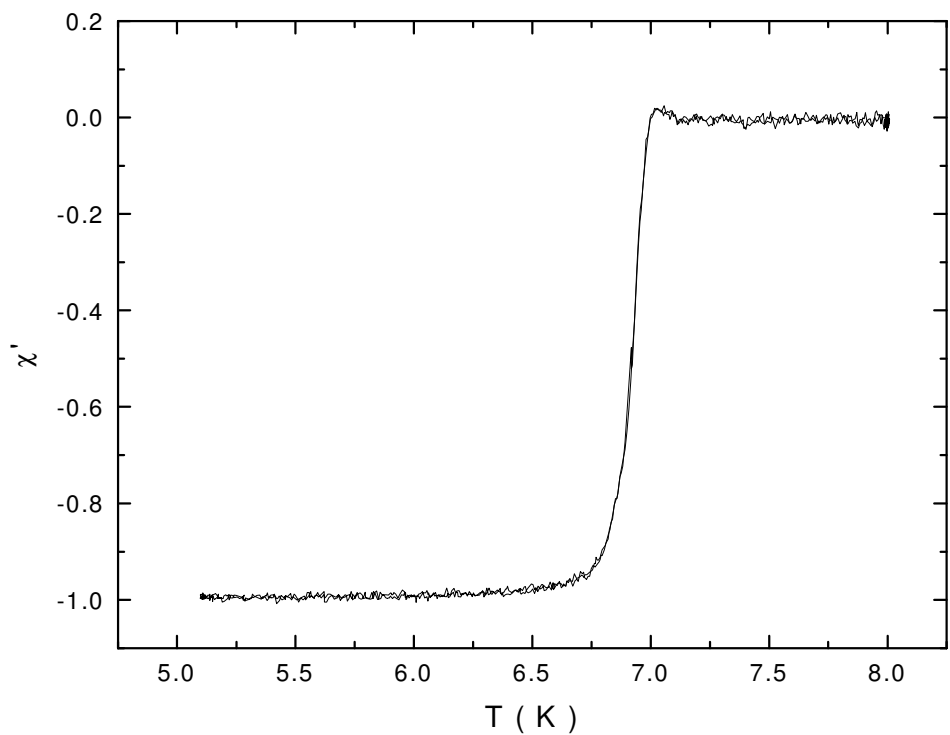

Fig.2 Ac-susceptibility superconducting transition in a Pb film prepared using a template from 900 nm polystyrene spheres (#900).

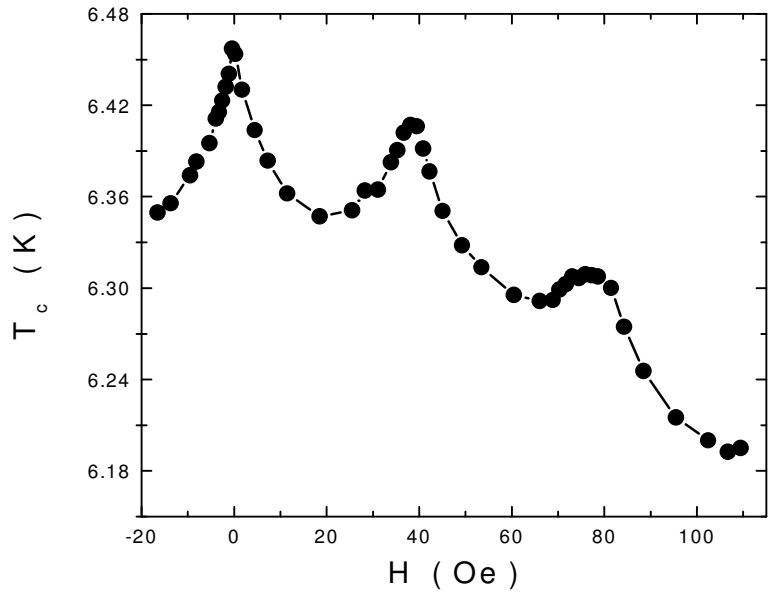

Fig.3 Little-Parks effect in a Pb film prepared using a template of 700 nm polystyrene spheres (#700b)

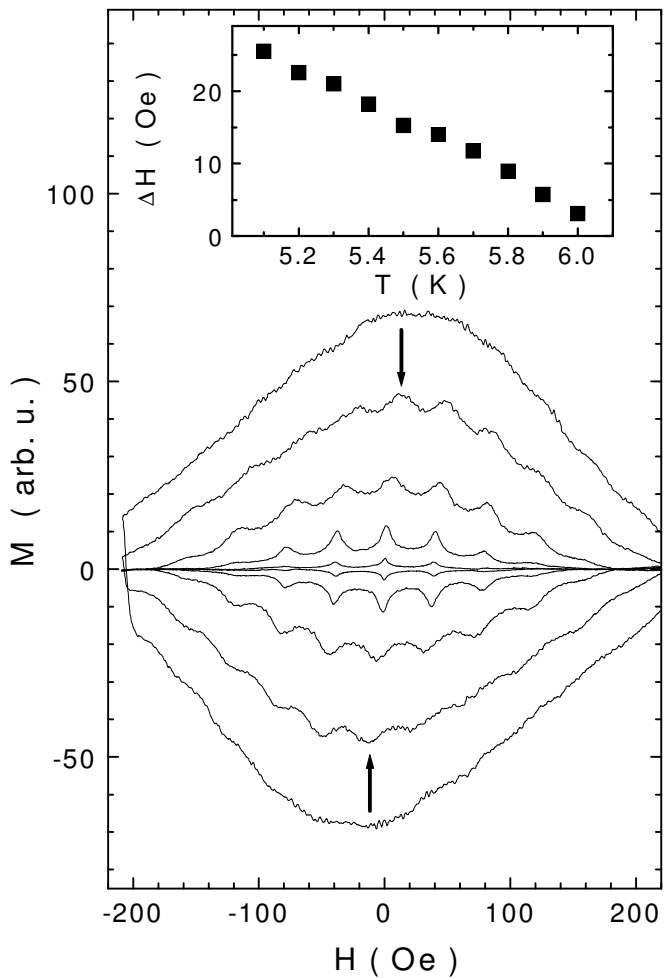

Fig.4 Magnetization curves at different temperatures (5.5K, 5.7K, 5.9K, 6.1K and 6.3K) for a Pb film prepared using a template from 700 nm polystyrene spheres (#700b). Arrows show position of n=0 maximum for 5.7K. The inset shows the temperature dependence of the switching field ΔH for H=200Oe.

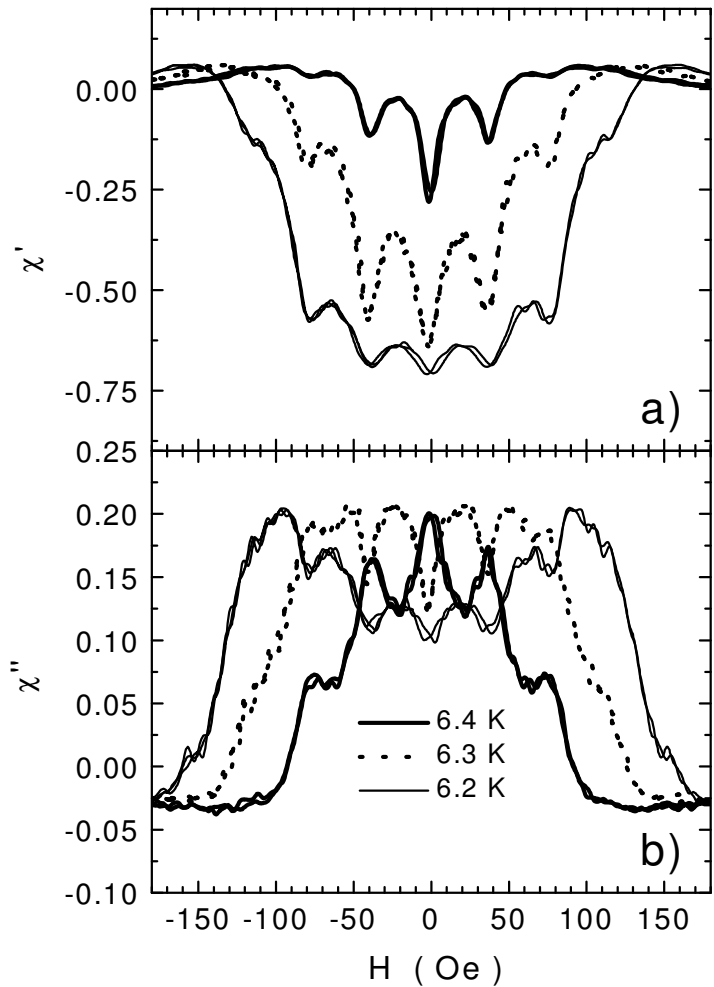

Fig.5 Magnetic field dependence for in-phase $\chi'$ (a) and out-of-phase $\chi''$ (b) components of ac-susceptibility ($H_{ac}$ = 0.15 Oe) for a Pb film prepared using 700 nm polystyrene spheres (#700b).

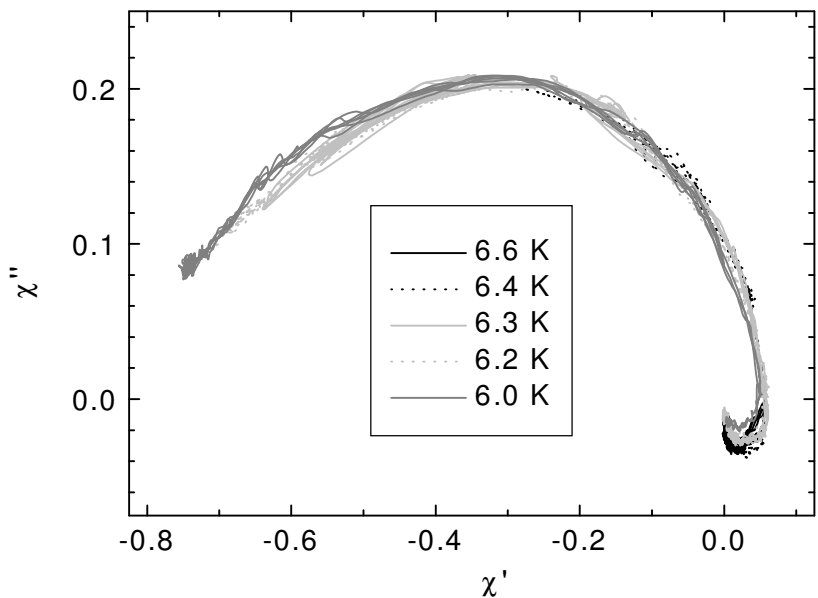

Fig.6 Cole-Cole plot of $\chi''$ vs. $\chi'$ at different temperatures.